\documentclass[journal]{IEEEtran}

\ifCLASSINFOpdf
\else
   \usepackage[dvips]{graphicx}
\fi
\usepackage{url}

\usepackage{graphicx}
\usepackage{xcolor}

\begin{document}

\title{Copy-Move Detection in Optical Microscopy: A Segmentation Network and A Dataset}

\author{Hao-Chiang Shao$^{\dagger}$, \IEEEmembership{Member, IEEE}, Yuan-Rong Liao, Tse-Yu Tseng, Yen-Liang Chuo, and Fong-Yi Lin
\thanks{Manuscript received on June 12, 2024. This work was supported
in part by 
NSTC 112-2221-E-005-080 and iGroup Asia-Pacifics. (Corresponding Author: Prof. H.-C. Shao; Email: shao.haochiang@gmail.com)}
\thanks{H.-C. Shao, Y.-L. Chuo, and F.-Y. Lin are with the Institute of Data Science and Information Computing, National Chung Hsing University, Taiwan. }
\thanks{Y.-R. Liao is with SOE Technology Inc., Taiwan.}
\thanks{T.-Y. Tseng is with iGroup (Asia-Pacific) Ltd., Taiwan.}
}

\markboth{IEEE Signal Processing Letters, Vol. XX, No. XX, MMMMMM 20YY}
{Shell \MakeLowercase{\textit{et al.}}: Bare Demo of IEEEtran.cls for IEEE Journals}
\maketitle

\begin{abstract}
With increasing revelations of academic fraud, detecting forged experimental images in the biomedical field has become a public concern. The challenge lies in the fact that copy-move targets can include background tissue, small foreground objects, or both, which may be out of the training domain and subject to unseen attacks, rendering standard object-detection-based approaches less effective. To address this, we reformulate the problem of detecting biomedical copy-move forgery regions as an intra-image co-saliency detection task and propose CMSeg-Net, a copy-move forgery segmentation network capable of identifying unseen duplicated areas. 
Built on a multi-resolution encoder-decoder architecture, CMSeg-Net incorporates self-correlation and correlation-assisted spatial-attention modules to detect intra-image regional similarities within feature tensors at each observation scale. This design helps distinguish even small copy-move targets in complex microscopic images from other similar objects. Furthermore, we created a copy-move forgery dataset of optical microscopic images, named FakeParaEgg, using open data from the ICIP 2022 Challenge to support CMSeg-Net’s development and verify its performance. \textcolor{blue}{Extensive experiments demonstrate that our approach outperforms previous state-of-the-art methods on the FakeParaEgg dataset and other open copy-move detection datasets, including \textbf{CASIA-CMFD}, \textbf{CoMoFoD}, and \textbf{CMF}. The FakeParaEgg dataset, our source code, and the CMF dataset with our manually defined segmentation ground truths} \textcolor{blue}{are} available at \textcolor{gray}{https://github.com/YoursEver/FakeParaEgg}.

\end{abstract}


\IEEEpeerreviewmaketitle

\section{Introduction}
\label{sec01:intro}

Biomedical image forgery issues have garnered public attention due to recent, ongoing revelations of academic fraud or misconduct \cite{MCCOOK2018Cancer,piller2022blots,kaiser2023should,thorp2024genuine}. Accusations of falsification emerging on platforms such as the ``PubPeer'' forum cast doubt even on esteemed scholars and previously influential research works. 
These cases typically involve illegal instances of image reuse or manipulation, often employing illicit post-processing approaches like deliberate cropping 
and copy-move manipulations \cite{shao2018unveiling}. Identifying such falsifications poses a significant challenge for publishers and researchers, 
and thus there is a growing public concern for methods capable of detecting fraudulent biomedical images.

\textcolor{blue}{
In real-world academic fraud cases, two common types of biomedical image manipulation are \textbf{copy-move} and \textbf{image reuse} \cite{bik2016prevalence}, the latter of which can be considered a form of copy-move forgery when the reused region appears alongside its source image.  
However, typical copy-move detection methods, often developed using an object detection backbone and natural image datasets, may struggle to identify duplications in biomedical images. This is because biomedical images frequently contain foreground objects with similar geometric shapes and uniform appearances, while their backgrounds often feature patterns resembling the foreground. Additionally, biomedical images involve texture patterns that are not adequately addressed by natural image datasets. As a result, few forgery detectors are specialized for the biomedical domain. }

To address these challenges, we created a microscopic copy-move image dataset named \textbf{FakeParaEgg} using the public data provided by the  \textit{ICIP 2022 Challenge: Parasitic Egg Detection and Classification in Microscopic Images}~\cite{ICIP2022Para,anantrasirichai2022icip}. Additionally, we redefined the copy-move detection task as an intra-image co-saliency detection problem, targeting duplicated regions within an image---whether foreground objects or background tissues. \textcolor{blue}{Accordingly, we devised a copy-move segmentation network (CMSeg-Net) capable of segmenting co-salient regions with out-of-domain unseen patterns by exploiting intra-image regional similarities in a multiresolution fashion. }

This paper thus makes three contributions. First, we reframe the copy-move forgery detection problem as an intra-image co-saliency detection issue and propose a lightweight segmentation model that is robust against \textcolor{blue}{
out-of-domain patterns even when exposed to unseen attacks such as background removal, compression, and scaling, which are commonly encountered in academic fraud investigations. }
Second, we introduce a correlation-assisted spatial-attention (CoSA) module to enhance the recognition of multiple intra-image duplicated regions, including small foreground objects and those similar to background areas, within the images. 
Third, we introduce \textcolor{blue}{and release} an optical microscopic copy-move forgery image dataset, named \textbf{FakeParaEgg}, with this paper to facilitate the future advancement of academic fraud detection techniques.

\section{Related Works}
\label{sec02:review}
Recent copy-move detection methods can be categorized into two groups: those relying on modern convolutional neural networks (CNN) and those employing conventional strategies such as feature-based methods and manual inspections for real-world academic misconduct investigations. Modern CNN-based approaches, 
typically address the copy-move detection problem by treating it as an object detection task. For example, Dense-InceptionNet \cite{zhong2019end} utilizes a single-branch dense-net based pyramid feature extractor to facilitate feature correlation matching for copy-move detection, while RGB-N \cite{zhou2018learning} and BusterNet \cite{wu2018busternet} employ a two-branch architecture for detecting similar objects based on matched features. In addition, some literature, such as Mantra-Net \cite{wu2019mantra}, views copy-move detection as a local anomaly detection issue and employs anomaly detection strategies to identify local forgeries. Another category of approaches tackles this problem using segmentation strategies centered around a feature affinity matrix, exemplified by RRU-Net \cite{bi2019rru} and DOA-GAN \cite{islam2020doa}. Recent methods of this kind, such as SelfDM-SA \cite{liu2021two}, CMFDFormer \cite{liu2023cmfdformer}, and UCM-Net \cite{weng2023ucm}, leverage the attention mechanism\textcolor{blue}{---widely used in computer vision applications, such as \cite{zhang2021semantic,zhang2023generative}---}to construct a sophisticated feature extraction backbone. This backbone enables the network model to provide a segmentation result with the improved assistance of intra-image correlation computations. However, these methods were primarily designed for natural images, and experiments indicate their inadequacy in handling real-world biomedical copy-move forgeries.

\textcolor{blue}{
In contrast, traditional feature-based methods identify copy-move regions by matching descriptors of high-level features. Examples include the dense-field approach by Cozzolino et al. \cite{cozzolino2015efficient}, Pun et al.'s regional feature point matching method \cite{pun2015image}, and the hierarchical feature point matching method by Li and Zhou \cite{li2018fast}. However, these methods face challenges when applied to images with weak edges or indistinct feature points, such as gel electrophoresis images or microscopic images containing semi-transparent foreground objects. Consequently, manual inspection remains a key tool in academic misconduct investigations, as highlighted by Piller \cite{piller2022blots}. This highlights the need for a new dataset and an improved copy-move detector. } 

\section{Method}
\label{sec03:method}

\begin{figure}[!t]
    \centering
    \includegraphics[width=0.49\textwidth]{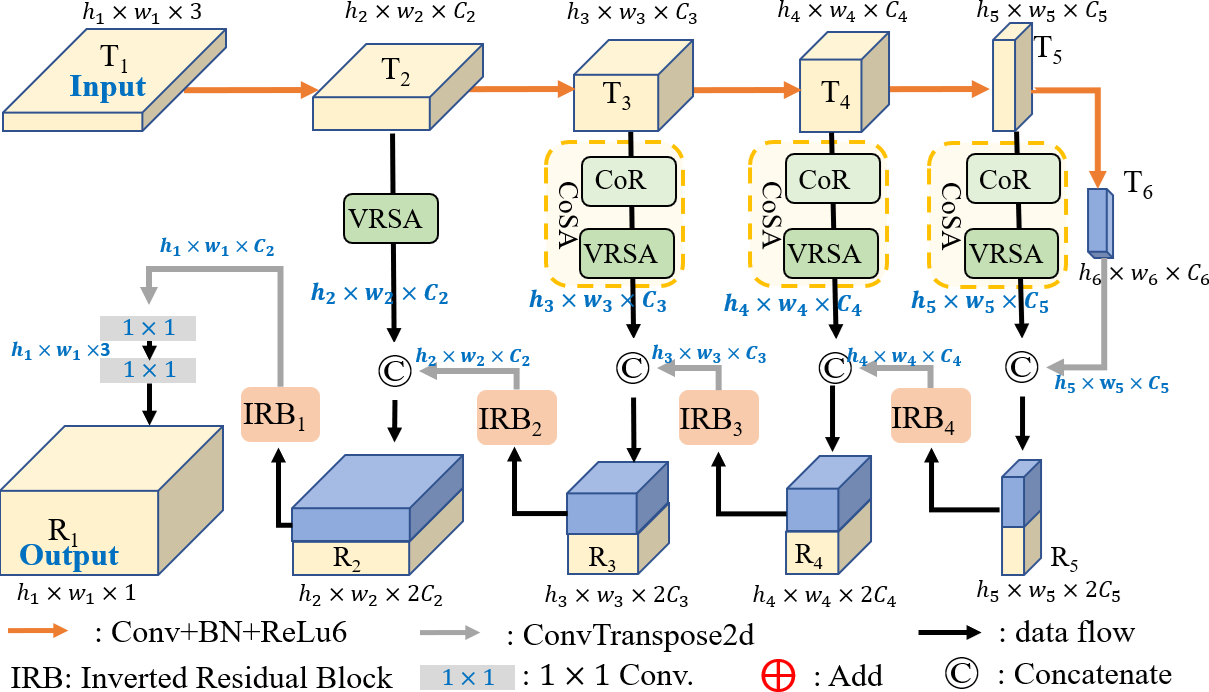}
    \caption{Architecture of the \textbf{CMSeg-Net}. 
    The encoding path from  $\mathbf{T}_1$ to $\mathbf{T}_6$ follows MobileNetV2, using ReLU6 activation for robustness, while the decoding path from $\mathbf{R}_5$ to $\mathbf{R}1$ utilizes the \textit{inverted residual block} (IRB) for reconstruction.  
    The VRSA block enhances its input with a spatial attention submodule featuring a variable-sized receptive field, while the CoR block derives spatial correlation for copy-move detection. Note that $C_2=16$, $C_3=24$, $C_4=32$, $C_5=96$, $C_6=1280$, and $(h_{i+1}, w_{i+1})=(\frac{1}{2}h_i,\frac{1}{2}w_i)$. 
    }
    \label{fig:archi}
\end{figure}

By reformulating the copy-move detection task as an intra-image co-saliency detection problem, we develope CMSeg-Net (Copy-move Segmentation Net), which aims to identify regions with highly correlated features. CMSeg-Net utilizes an encoder-decoder structure, commonly adopted in previous cosaliency detection networks \cite{zhang2020gradient,zhang2021deepacg}, with a MobileNet-v2 \cite{sandler2018mobilenetv2} encoder for feature representability and lightweight purposes. Additionally, to derive inter-pixel feature correlation for detecting duplicated copy-move regions, we designe a correlation-assisted spatial-attention (CoSA) module. This module consists of a correlation submodule (CoR) and a spatial attention submodule with a variable-sized receptive field (VRSA), implemented by incorporating the spatial attention module 
\cite{woo2018cbam} 
with a variable receptive field adopted in the atrous spatial pyramid pooling 
layer \cite{chen2017deeplab}. Finally, to \textcolor{blue}{efficiently} re-organize the feature tensors enhanced by CoSA modules, we adopt the inverted residual block (IRB) to accomplish the decoding path of CMSeg-Net, \textcolor{blue}{reducing the memory requirement \cite{sandler2018mobilenetv2}}. 
The framework is depicted in Fig. \ref{fig:archi}

In the CMSeg-Net design, a MobileNet-v2 pretrained on ImageNet \cite{deng2009imagenet} serves as the initial encoder for the learning process. This choice is motivated by real-world copy-move scenarios in which any object could be a potential forgery target, necessitating a feature extraction backbone capable of deriving intricate features from complex scenes. Based on the feature tensors extracted by the MobileNet-v2 encoder, CMSeg-Net employs a CoSA module to evaluate intra-image feature correlations and identify highly correlated spatial positions within the mid-level feature tensors. Subsequently, the submodule refines the feature tensor based on the contribution from the \textit{i}-th pixel to the \textit{j}-th pixel, guaranteeing accurate detection and segmentation of copy-move areas. 
Note that only the feature tensors of the three intermediate levels, \textit{i.e.}, $\mathbf{T}_3$, $\mathbf{T}_4$, and $\mathbf{T}_5$, are processed by the full CoSA module, while the CoSA module is simplified for the $\mathbf{T}_2$ tensor and neglected for the $\mathbf{T}_6$ tensor. 
This is because the deepest/topmost feature captures too global/local information, which is not effective in identifying copy-move areas.  
The details of our CMSeg-Net design will be elaborated in the following subsections. 
\noindent $\bullet $\textbf{ Correlation Submodule (CoR):}
The CoR module is designed to capture the intra-image similarity of a copy-move forgery image. Given a feature tensor $\mathbf{T}_i$ with size $h_i\times w_i \times C_i$, the CoR module first derives an $(h_iw_i)\times(h_iw_i)$ initial affinity matrix $\hat{\mathbf{A}}_i$ as 
$\hat{\mathbf{A}}_i = \mathbf{F}_i^T \mathbf{F}_i$, 
and 
$\mathbf{F}_i$ is a $C_i\times(h_iw_i)$ reshaped version of $\mathbf{T}_i$ with its column vectors normalized.  Clearly, $\hat{\mathbf{A}}_i(\mathbf{s},\mathbf{t})$ records the cosine similarity between the feature vectors of the $\mathbf{s}$-th and the $\mathbf{t}$-th positions. 
\textcolor{blue}{
Then, we obtain the affinity matrix $\mathbf{A}_i= \hat{\mathbf{A}}_i \odot \mathbf{\Phi}_i$ by suppressing the self-similarity around the main diagonal of $\hat{\mathbf{A}}_i$ with a suppression matrix $\mathbf{\Phi}_i$. 
Here,} $\mathbf{\Phi}_i(\mathbf{s}, \mathbf{t}) = 1-e^{-\frac{(s_x-t_x)^2+(s_y-t_y)^2}{2\gamma^2}}$ with $\gamma$ typically ranging between 3 and 5 by default; $(s_x, s_y)$ and $(t_x, t_y)$ are respectively the coordinates of $\mathbf{s}$ and $\mathbf{t}$ on the $h_i\times w_i$ domain; and, $\odot$ denotes the Hadamard product. 
Because the second term of $\Phi_i$ is a radial basis function, $\Phi_i(\mathbf{s},\mathbf{t})$ approaches 0 when $\mathbf{s}$'s coordinate approaches $\mathbf{t}$'s, and therefore the self-similarity around the main diagonal of  $\hat{\mathbf{A}}_i$ can be suppressed. Finally, we reshape $\mathbf{A}_i$ back to a $(h_iw_i)\times h_i\times w_i$ tensor, and then follow Liu et al.'s \cite{liu2021two} strategy to output a $k \times h_i\times w_i$ tensor $\mathbf{K}$ by selecting the $k$ largest values of each $(h_iw_i)\times 1\times 1$ vector in descending order. 

\noindent $\bullet $\textbf{Spatial attention with  variable-sized receptive field (VRSA):}
Our VRSA submodule consists of an atrous spatial pyramid pooling (ASPP) block \cite{chen2017deeplab} and a spatial attention (SAM) block used in CBAM \cite{woo2018cbam}. 
\textcolor{blue}{In CMSeg-Net, the ASPP block is implemented with four $3\times3$ convolutional layers using dilation rates of \{4, 8, 12, 16\} to extract contextual features from the tensor $\mathbf{K}$ at different fields of view. }
The resulting four feature maps are concatenated and then fed into a $1\times1$ convolution layer to derive the tensor $\tilde{\mathbf{K}}_i$  that captures local and image-level features. Finally, this tensor is further fed into a spatial attention block to yield 
    $\mathbf{B}_i = \tilde{\mathbf{K}}_i + \mathrm{M}_s(\tilde{\mathbf{K}}_i)$. 
Here, 
$\mathrm{M}_s(\cdot)\in \mathrm{R}^{H\times W}$ is the spatial attention map defined as 
    $\mathrm{M}_s(\tilde{\mathbf{K}}_i) = \sigma( f^{7}(\tilde{\mathbf{K}}_i) )$, 
where $\sigma$ and $f^7$  denote the sigmoid function and a $7\times7$ convolution layer, respectively. 

\noindent $\bullet $\textbf{Decoding Path}: 
CMSeg-Net's decoding path structurally mirrors the multi-resolution reconstruction framework used in wavelet and pyramid analysis \cite{shao2011optimal,shao2024rgbt2hs}.
This design leads to the reconstruction result 
    $\mathrm{R}_i = 
    \mathrm{B}_{i}
    \mbox{\large{ \textcircled{c}}} \, f^d_i\big(\, \mathrm{IRB}_i(\mathrm{R}_{i+1} )\,\big)$  for 
    $2\leq i\leq4$\mbox{.}
Here, $f^d_i(\cdot)$ denotes the \textbf{ConvTranspose2D} routine for the tensor derived by the \textit{inverted residual block}~\cite{chen2017deeplab} at the \textit{i}-th scale 
in the decoding path%
, and $\mathrm{B}_i = \mathrm{CoSA}(\mathrm{T}_{i})$. 
Note that in order to fit the hardware limitation, the CoSA module for $\mathrm{T}_2$ consists of only a VRSA submodule, and the CoR submodule is omitted. 
Additionally, the CoSA module for $\mathrm{T}_6$ is omitted because $\mathrm{T}_6$ has too coarse a spatial resolution to capture inter-region spatial correlations. 

\noindent $\bullet $\textbf{ Total Loss}: As a segmentation network, CMSeg-Net is trained using binary cross-entropy loss and Dice loss. 


\section{Experiment Result}
\label{sec04:experiment}
\subsection{Datasets}

\textcolor{blue}{We evaluate the model performance on the  \textbf{FakeParaEgg}, \textbf{CASIA-CMFD}, \textbf{CoMoFoD} \cite{tralic2013comofod}, and \textbf{CMF} \cite{warif2017sift} datasets.} 

\textcolor{blue}{\textbf{FakeParaEgg} is our self-made microscopic image dataset designed to simulate copy-move manipulations commonly encountered in real-world academic fraud cases. Unlike existing open datasets, FakeParaEgg consists of forgery images where foreground objects have had their backgrounds removed (i.e., made transparent), making these copy-moves difficult to detect using traditional methods due to the nonidentical background context.  Moreover, \textbf{FakeParaEgg} introduces a wide range of complexities, including low-contrast dark backgrounds, high-contrast bright backgrounds, nearly transparent foreground objects, and both simple and complex backgrounds. It also features varied foreground sizes, making it a highly diverse resource for developing robust detection methods for academic fraud investigations. 
%
%
\textbf{FakeParaEgg} contains 1,400 forged images for training and 100 for testing. To create it, we cropped parasitic eggs using ground-truth bounding boxes from  \cite{ICIP2022Para}, removed the backgrounds using \textbf{Rembg} \cite{rembg}, and randomly pasted the background-free patches onto the original image, producing the final synthetic copy-move forgeries.} 

\textcolor{blue}{
Additionally, 
\textbf{CASIA-CMFD}, a subset of \textbf{CASIA~V2.0} \cite{dong2013casia}, contains 1,309 copy-move images and is widely used for performance evaluation in copy-move detection studies. \textbf{CoMoFoD} and \textbf{CMF} are used to assess the model's generalizability on out-of-domain (OOD) data with unseen attacks. \textbf{CoMoFoD} includes 25 categories, covering geometrical attacks (F), brightness changes (BC), contrast adjustments (CA), color reduction (CR), image blurring (IB), JPEG compression (JC), and noise addition (NA), with varying levels of attack intensity, each consisting of 200 attacked copy-move samples. In contrast, \textbf{CMF} has 17 categories, including no attack (Non), scaling with rotation (SR), mirroring (MIR), rotation (Ro), and scaling (Sc), each with 15 samples. 
}

\subsection{Result and Discussion}

\begin{figure}[!t]
    \centering
    \includegraphics[width=0.49\textwidth]{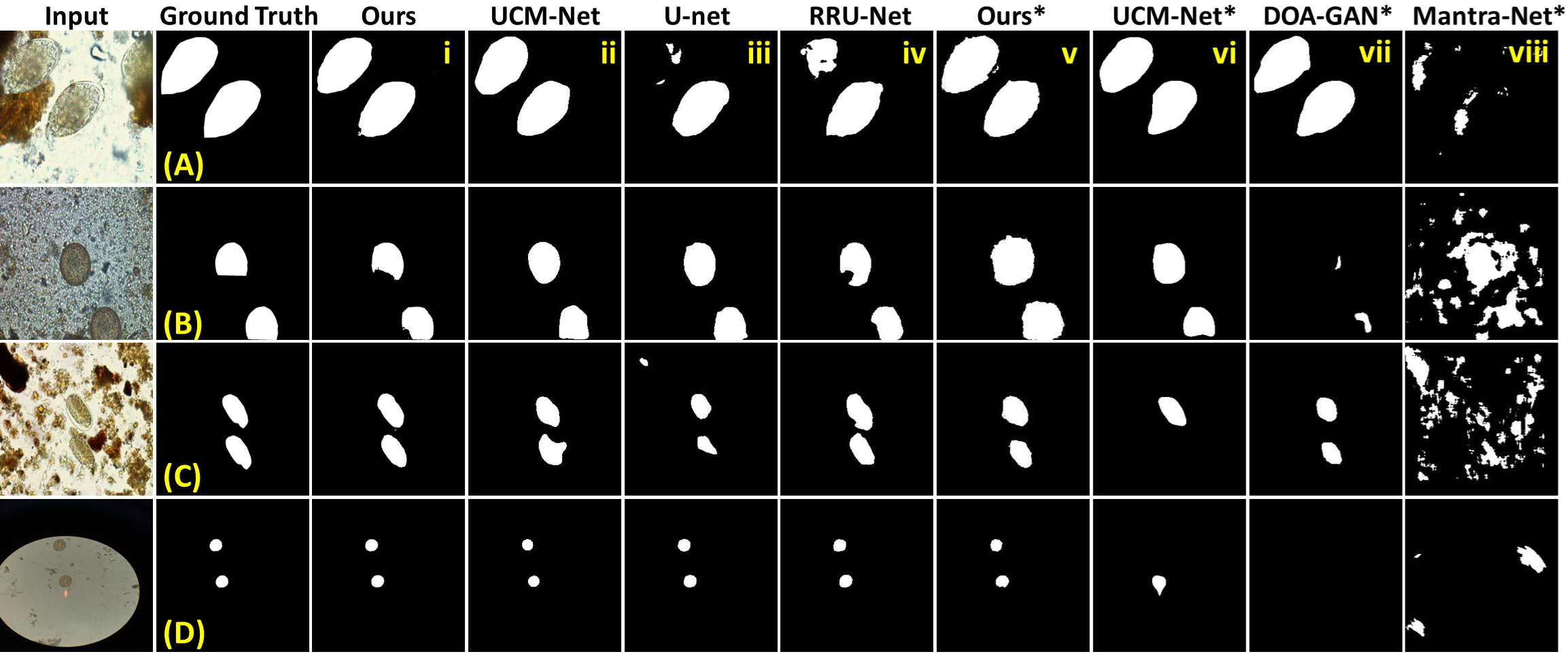}
    \caption{Visual performance comparison among methods on \textbf{FakeParaEgg}. Note that Columns \textbf{iv}-\textbf{vii} are out-of-domain generalizability tests.}
    \label{fig:paraeggresult}
\end{figure}

\begin{table}[!t]
    \centering
    \caption{Performance comparison on \textbf{FakeParaEgg}. (A) single domain tests, (B) \textcolor{blue}{out-of-domain generalizability tests} }    
    \scriptsize
    \begin{tabular}{|c|l|c|c|c|c|c|}
    \hline
        & Method & mF1 & mIoU  & Prec. & Recall & Spec. \\
    \hline
        & U-Net \cite{ronneberger2015u} & .749 & .616  & .770 & .855 & .9934 \\
    (A) & RRU-Net \cite{bi2019rru} & .777 & .652 & .788 & .898 & .9938\\
        & UCM-Net \cite{weng2023ucm} & .713 & .586 & .782 & .867 & .9937\\
        
        & Ours & \textcolor{blue}{\textbf{.922}} & \textcolor{blue}{\textbf{.885}} & .878 & .970 & .9972\\ \hline \hline
        
        & {ManTra-Net} \cite{wu2019mantra} & .099 & .060 & .094 & .280 & .9302\\
        & {DOA-GAN} \cite{islam2020doa} & .390 & .315 & .792 & .603 & .9959\\
    (B) & {UCM-Net} \cite{weng2023ucm} & .478 & .396 & .714 & .742 & .9923\\
        & Ours  & \textcolor{blue}{\textbf{.739}} & \textcolor{blue}{\textbf{.582}} & .828 & .667 & .9962\\
    \hline
    \end{tabular}
    \label{tab:praraegg_performance}
\end{table}

\noindent \textbf{\underline{FakeParaEgg}:}
Fig. \ref{fig:paraeggresult} \textcolor{blue}{(Columns i-iv)} demonstrates the copy-move detection results on samples from the \textbf{FakeParaEgg} dataset. These experimental results confirm that our method effectively detects and segments copy-move regions in intricate microscopic scenes. Even in scenes with small, similar objects or complex backgrounds, our method successfully identifies forged areas. Table \ref{tab:praraegg_performance} summarizes the performance metrics of several copy-move detection schemes on the \textbf{FakeParaEgg} dataset. Our CMSeg-Net outperforms previous approaches 
in terms of mean F1 and mean IoU values.  
Note that the U-net here serves as a performance baseline, trained using copy-move images and their corresponding ground-truth segmentation masks with binary cross-entropy loss. 
%
Additionally, Fig. \ref{fig:paraeggresult} \textcolor{blue}{(Columns v-viii)} and Table \ref{tab:praraegg_performance}(B) \textcolor{blue}{present a comparison of out-of-domain generalizability.} 
In this experiment, our CMSeg-Net was trained on CASIA-CMFD and tested on FakeParaEgg, while the test results for the other three methods were derived from their officially released best  models.  
\textcolor{blue}{This experiment demonstrates CMSeg-Net's robustness on bio-images outside the training domain.}

%

\noindent \textbf{\underline{CASIA-CMFD}:}
\begin{figure}[!t]
    \centering
    \includegraphics[width=0.48\textwidth]{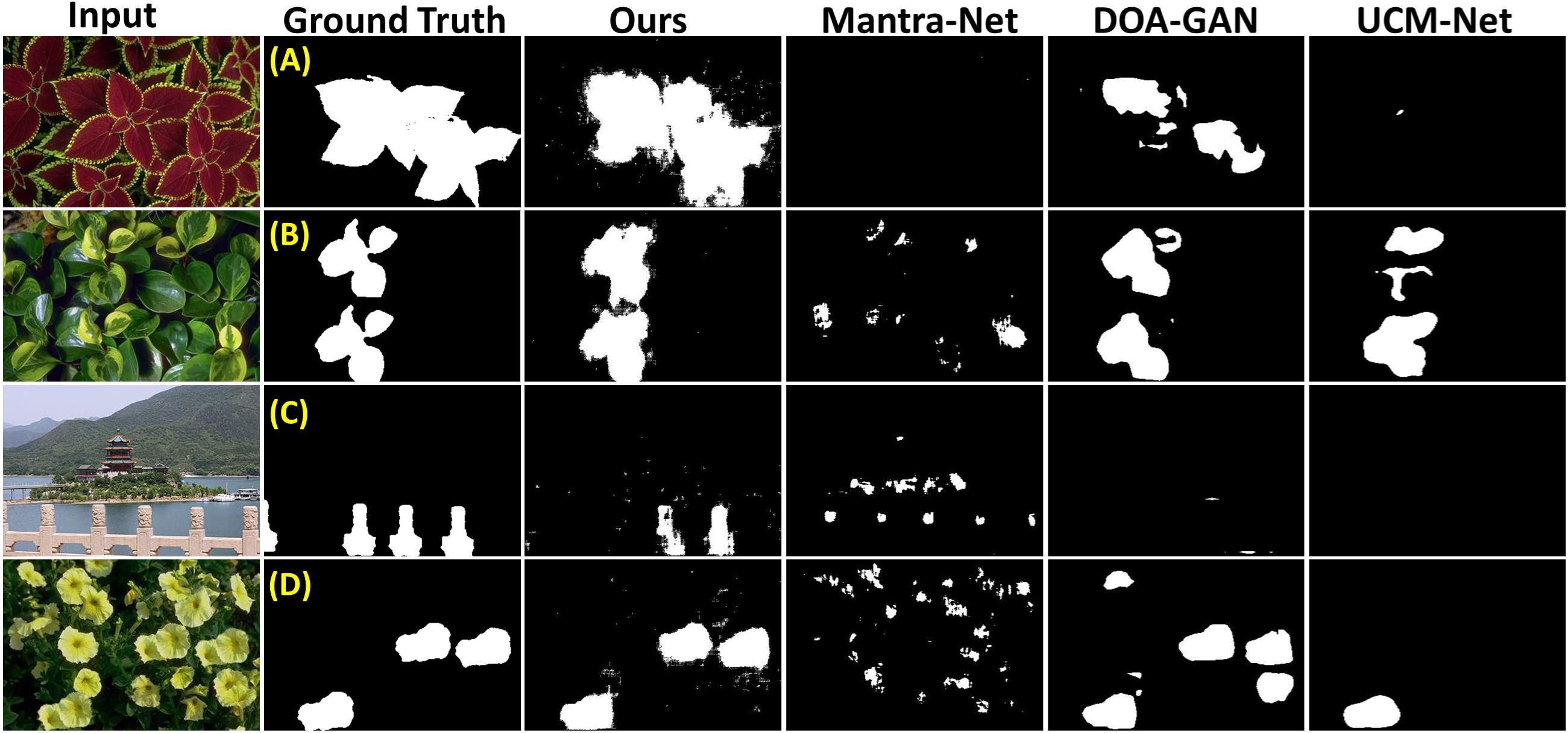}
    \caption{Visual performance comparison of our method against ManTra-Net, DOA-GAN, and UCM-Net on the \textbf{CASIA-CMFD} dataset.}
    \label{fig:resultcasia}
\end{figure}
\begin{table}[!t]
    \centering
    \caption{\textcolor{blue}{Comparison of model performance on \textbf{CASIA-CMFD} }}    
    \scriptsize
    \begin{tabular}{|l|c|c|c|c|c|c|}
    \hline
       Method & mF1 & Prec. & Recall & Spec. & Time(ms)& \#Para.(M)\\
       \hline
       Busternet \cite{wu2018busternet} &  .456  & .524 & .692 & .942 & 10.91 & 15.53\\
       RGB-N \cite{zhou2018learning} & .408  & -- & -- & -- & -- &--\\
       RRU-Net \cite{bi2019rru} & .212 & .267 & .255 & .921 & 13.41 & 4.10\\
       ManTra-Net \cite{wu2019mantra} & .060  & .450 & .042 & .994 & 46.93 & 3.81\\
       DenseIncept. \cite{zhong2019end} & .643  & .709 & .589 & -- & -- & --\\
       DOA-GAN \cite{islam2020doa} & .462  & .790 & .505 & .985 & 8.60 & 26.97\\
       AR-Net \cite{zhu2020ar} & .455  & .583 & .373 & -- & -- &--\\
       SD-Net \cite{li2022image} & .481  & .575 & .513 & -- & -- & --\\
       UCM-Net \cite{weng2023ucm} & .543  & .792 & .627 & .982 & 23.37 &11.14\\
       \hline
       \hline
       Ours  & \textcolor{blue}{.643} & .887 & .505 & .992 & 12.13 &5.16\\
    \hline
    \end{tabular}
    \label{tab:performanceCASIA}
\end{table}
\begin{table}[!t]
    \centering
    \caption{\textcolor{blue}{Generalizability tests on OOD data with unseen attacks. 
    The numbers indicate the counts of detected masks with F1$>0.5$.} }   
    \tiny
    \begin{tabular}{|c||c|c|c|c|c|c|}
        \hline
        \multicolumn{7}{|c|}{(A) CoMoFoD \cite{tralic2013comofod}}\\ \hline
        \multicolumn{1}{|c|}{Attack} & 
        \multicolumn{1}{c|}{BusterNet } &
        \multicolumn{1}{c|}{RRU-Net } &
        \multicolumn{1}{c|}{ManTra-Net } &
        \multicolumn{1}{c|}{DOA-GAN } &
        \multicolumn{1}{c|}{UCM-Net } &
        \multicolumn{1}{c|}{Ours} \\
        \multicolumn{1}{|c|}{} &
        \multicolumn{1}{c|}{2018 \cite{wu2018busternet}} &
        \multicolumn{1}{c|}{2019 \cite{bi2019rru}} &
        \multicolumn{1}{c|}{2019 \cite{wu2019mantra}} &
        \multicolumn{1}{c|}{2020 \cite{islam2020doa}} &
        \multicolumn{1}{c|}{2024 \cite{weng2023ucm}} &
        \multicolumn{1}{c|}{} \\ \hline
        F& 
            117& 25& 13 & 
            102& 89& 112\\ \hline
        BC1& 
            116& 25& 10 & 
            104& 88& 110\\ \hline
        BC2& 
            115& 23& 9 & 
            104& 90& 110\\ \hline
        BC3& 
            109& 24 & 9 & 
            100& 93& 109\\ \hline
        CA1& 
            117& 25& 10 & 
            103& 89& 116\\ \hline
        CA2& 
            116& 27& 11 & 
            103& 92& 117\\ \hline
        CA3& 
            116& 26& 14 & 
            103& 94& 117\\ \hline
        CR1& 
            117& 25& 13 & 
            103& 89& 113\\ \hline
        CR2& 
            116& 25& 13 & 
            103& 88& 114\\ \hline
        CR3& 
            116& 23& 12 & 
            102& 88& 114\\ \hline
        IB1& 
            113& 27& 6 & 
            93& 91& 112\\ \hline
        IB2& 
            98& 25& 3 & 
            80& 102& 123\\ \hline
        IB3& 
            93& 20& 3 & 
            72& 84& 117\\ \hline
        JC1& 
            60& 25& 7 & 
            77& 85& 94\\ \hline
        JC2& 
            77& 29& 14 & 
            82& 87& 107\\ \hline
        JC3& 
            86& 26& 8 & 
            91& 87& 111\\ \hline
        JC4& 
            103& 26& 9 & 
            94& 84& 117\\ \hline
        JC5& 
            99& 25& 11 & 
            95& 89& 118\\ \hline
        JC6& 
            101& 25& 12 & 
            94& 87& 115\\ \hline
        JC7& 
            107& 22& 7 & 
            97& 87& 116\\ \hline
        JC8& 
            109& 23& 10 & 
            100& 91& 116\\ \hline
        JC9& 
            106& 28& 10 & 
            97& 85& 114\\ \hline
        NA1& 
            100& 26& 3 & 
            56& 82& 67\\ \hline
        NA2& 
            102& 24& 8 & 
            62& 77& 86\\ \hline
        NA3& 
            -& 21& 21 & 
            95& 87& 118\\ \hline
        \textbf{Total}& 
            \textbf{2509}& \textbf{620}& \textbf{246} & 
            \textbf{2312}& \textbf{2205}& \textcolor{blue}{\textbf{2763}}\\ \hline

        \hline
        \multicolumn{7}{|c|}{(B) CMF \cite{warif2017sift}}\\ \hline
        \multicolumn{1}{|c|}{Attack} & 
        \multicolumn{1}{c|}{BusterNet } &
        \multicolumn{1}{c|}{RRU-Net } &
        \multicolumn{1}{c|}{ManTra-Net } &
        \multicolumn{1}{c|}{DOA-GAN } &
        \multicolumn{1}{c|}{UCM-Net } &
        \multicolumn{1}{c|}{Ours} \\ \hline
        Non& 
            14& 5& 2 & 
            14& 15& 14\\ \hline
        SR1& 
            13& 6& 2 & 
            14& 14& 15\\ \hline
        SR2& 
            13& 4& 3 & 
            14& 15& 15\\ \hline
        SR3& 
            10& 6 & 1 & 
            11& 13& 14\\ \hline
        SR4& 
             1& 5 & 1 & 
             1& 8 & 7\\ \hline
        SR5& 
             2& 5 & 2 & 
             1& 6 & 8 \\ \hline
        MIR& 
             3& 3 & 2 & 
             1& 11& 15\\ \hline
        Ro1& 
            13& 5 & 1 & 
            14& 15& 15\\ \hline
        Ro2& 
             3& 6 & 1 & 
             4& 9 & 9\\ \hline
        Ro3& 
             2& 5 & 2 & 
             0& 2 & 4\\ \hline
        Ro4& 
             0& 4 & 2 & 
             0& 1 & 2\\ \hline
        Ro5& 
             0& 5 & 1 & 
             0& 0 & 2\\ \hline
        Sc1& 
            10& 3 & 0 & 
            10& 12& 14\\ \hline
        Sc2& 
            13& 5 & 0 & 
            14& 15& 15\\ \hline
        Sc3& 
            14& 4 & 3 & 
            14& 15& 15\\ \hline
        Sc4& 
            13& 4 & 4 & 
            14& 15& 14\\ \hline
        Sc5& 
            12& 5 & 4 & 
            11& 14& 12\\ \hline
        \textbf{Total}& 
            \textbf{136}& \textbf{80}& \textbf{31} & 
            \textbf{137}& \textbf{180}& \textcolor{blue}{\textbf{190}}\\ \hline
    \end{tabular}
    \label{tab:comofod_cmf}
\end{table}
Table \ref{tab:performanceCASIA} and Fig. \ref{fig:resultcasia}  present a performance comparison between our CMSeg-Net and previous state-of-the-art methods on CASIA-CMFD. 
This comparison further underscores the effectiveness of CMSeg-Net in general copy-move forgery detection tasks. Also, CMSeg-Net performs reliably even in scenarios of \textcolor{blue}{(i) copy-move of background textures (see Fig. \ref{fig:resultcasia}(A) and (B)), and (ii) multiple copy-move objects (see Fig. \ref{fig:resultcasia}(C) and (D)).}


\noindent
\textcolor{blue}{
\underline{\textbf{CoMoFoD} \& \textbf{CMF}}\textbf{:} Table \ref{tab:comofod_cmf} demonstrates CMSeg-Net's generalizability against out-of-domain (OOD) copy-move images with unseen attacks. All models were trained on CASIA-CMFD, and an attacked copy-move image was considered successfully detected if its detected mask achieved $F1>0.5$. This experiment evidences that CMSeg-Net is the most robust against OOD data with attacks such as brightness changes (BC), contrast adjustments (CA), JPEG compression (JC), scaling with rotation (SR), and mirroring (MIR), which are commonly used in real-world image forgery cases.
}

\textcolor{blue}{Morevoer, because CMSeg-Net  detects areas of high co-saliency within an image as potential copy-move regions, discrepancies may arise between the detected areas and the groundtruth masks when co-saliency is insufficiently strong due to features from neighboring regions, as demonstrated in Fig. \ref{fig:paraeggresult} and Fig.  \ref{fig:resultcasia}. Additionally, CMSeg-Net's high precision and high specificity, as shown in Tables \ref{tab:praraegg_performance} and \ref{tab:performanceCASIA}, highlight its ability to reduce false positives while maintaining accuracy, even in complex scenarios. Notably, Table  \ref{tab:performanceCASIA} also includes the computational time and parameter counts for several models. With a smaller parameter count and superior performance, CMSeg-Net proves to be an efficient and effective solution for copy-move forgery detection.}

\subsection{Ablation Study}

Table \ref{tab:ablationnew} presents the ablation study. The baseline model in the first row is a vanilla encoder-decoder network with a pretrained MobileNet-v2 encoder. The results indicate that the VRSA module (consisting of ASPP and SAM) contributes the most to the model's performance. Other components, such as the CoR module with $\Phi_i$ and the IRB blocks in the decoding path, are also critical to the CMSeg-Net design.   

\begin{table}[!t]
    \centering
    \caption{Ablation Study on CASIA-CMFD Dataset}   
    \scriptsize
    \begin{tabular}{|c|cc|c||c|c|c|c|c|}
    \hline
    CoR & \multicolumn{2}{c|}{VRSA} &  IRB & mF1 & mIoU & Prec.  & Recall & Spec.\\ \cline{2-3}
        & ASPP & SAM &      &     &     &      &   & \\
    \hline
       --  & -- & -- & --& .174 & .020 & .555 & .181 & .980\\ 
       --  & v & v & v & .261 & .147 & .506 & .176 & .979\\ 
       v  & -- & -- & v & .169 & .127 & .496 & .180 & .979 \\ 
       v  & v & -- & v & .598 & .407 & .904 & .447 & .994\\ 
       v  & -- & v & v & .268 & .129 & .515 & .181 & .979\\
       v  & v & v & -- & .475 & .286 & .877 & .326 & .994\\
    \hline
       v & v & v & v & \textcolor{blue}{.643} & \textcolor{blue}{.437} & \textcolor{blue}{.887} & \textcolor{blue}{.505} & \textcolor{blue}{.992} \\ 
    \hline
    \end{tabular}
    \label{tab:ablationnew}
\end{table}


\section{Concluding Remarks}
\label{sec05:conclu}
\textcolor{blue}{
In this paper, we introduced CMSeg-Net for identifying within-image copy-move regions based on intra-image co-saliency. CMSeg-Net builds upon the encoder-decoder framework widely used in co-saliency detection methods and incorporates correlation-assisted spatial attention to explore intra-image regional correlations. This architecture enables CMSeg-Net to detect copy-move objects in a multi-resolution fashion, ensuring that even small and previously unseen forgery objects in complex scenes are accurately identified. Comprehensive experiments on various datasets, including CASIA-CMFD, CoMoFoD, CMF, and FakeParaEgg, 
demonstrate CMSeg-Net's effectiveness and its generalizability against out-of-domain data with unseen attacks. }

\bibliographystyle{IEEEtran}
\bibliography{sec99_ref}

\end{document}